\documentclass[prd,preprint,superscriptaddress,showpacs,byrevtex]{revtex4}
\usepackage{bm}
\def\fsl#1{\setbox0=\hbox{$#1$}           % set a box for #1
   \dimen0=\wd0                                 % and get its size
   \setbox1=\hbox{/} \dimen1=\wd1               % get size of /
   \ifdim\dimen0>\dimen1                        % #1 is bigger
      \rlap{\hbox to \dimen0{\hfil/\hfil}}      % so center / in box
      #1                                        % and print #1
   \else                                        % / is bigger
      \rlap{\hbox to \dimen1{\hfil$#1$\hfil}}   % so center #1
      /                                         % and print /
   \fi}                                         %
\newcommand{\be}{\begin{equation}}
\newcommand{\ee}{\end{equation}}
\newcommand{\bea}{\begin{eqnarray}}
\newcommand{\eea}{\end{eqnarray}}
\newcommand{\beq}{\begin{equation}}
\newcommand{\eeq}{\end{equation}}
\newcommand{\beqs}{\begin{eqnarray}}
\newcommand{\eeqs}{\end{eqnarray}}

\begin{document}
\title{ There Is No Proof of Thermalized Quark-Gluon Plasma at RHIC and LHC }
\author{Gouranga C Nayak }\thanks{E-Mail: nayakg138@gmail.com}\thanks{G. C. Nayak was affiliated with C. N. Yang Institute for Theoretical Physics in 2004-2007.}
\affiliation{ C. N. Yang Institute for Theoretical Physics, Stony Brook University, Stony Brook NY, 11794-3840 USA}
\date{\today}
\begin{abstract}
Although Tevatron has discovered top quark and LHC (pp collisions) has discovered Higgs boson but the RHIC and LHC heavy-ion colliders have not discovered thermalized quark-gluon plasma (QGP). This is because the experimental data of top quark at Tevatron and the experimental data of Higgs boson at LHC are compared with the exact first principle calculation but the experimental data at RHIC and LHC are compared with simplistic models and assumptions which are not exact first principle calculation. In this paper we show that the exact first principle method to study quark-gluon plasma at RHIC and LHC is the nonequilibrium-nonperturbative QCD by using closed-time path integral formalism. Hence in the absence of such exact first principle calculation we conclude that there is no proof of thermalized quark-gluon plasma at RHIC and LHC.
\end{abstract}
\maketitle
\pagestyle{plain}
\pagenumbering{arabic}
\section{Introduction}

Just after $\sim 10^{-12}$ seconds of the big-bang our universe was filled with a hot and dense state of matter known as the quark-gluon plasma (QGP). Hence it is important to recreate this early universe matter (QGP) in the laboratory. The temperature of the quark-gluon plasma is $\gtrsim$ 200 MeV ($\gtrsim 10^{12}$ kelvin) which is much larger than the temperature of the stars (including the sun). The quark-gluon plasma is also the densest state of matter known in the universe (apart from the black hole). Due to the rapid expansion of the early universe, just after $\sim 10^{-6}$ seconds of the big-bang, the temperature of the quark-gluon plasma dropped so that protons and neutrons were formed for the first time from the quarks and gluons.

The relativistic heavy-ion colliders (RHIC) at BNL and the large hadron colliders (LHC) at CERN are the two experiments in the laboratory to recreate this early universe matter (the quark-gluon plasma) \cite{rn1,rn2}. Since the temperature $T$ and energy density $\epsilon$ are related by $\epsilon \propto T^4$ one finds that at the energy density $\gtrsim 2 ~{\rm GeV}/{\rm fm}^3$ the normal hadronic matter undergoes a phase transition to the quark-gluon plasma. Note that the energy density of the normal hadronic matter (heavy nucleus) is $\sim 01.5 ~{\rm GeV}/{\rm fm}^3$.

The energy density $\gtrsim 2 ~{\rm GeV}/{\rm fm}^3$ required to produce the quark-gluon plasma in the laboratory can be achieved by colliding two heavy nuclei at very high energy. The center of mass energy per nucleon in Au-Au heavy-ion collisions at RHIC is ${\sqrt s}_{NN}$ = 200 GeV and the center of mass energy per nucleon in Pb-Pb heavy-ion collisions at LHC is ${\sqrt s}_{NN}$ = 5.02 TeV. Since the center of mass energies per nucleon are very high at RHIC and LHC heavy-ion colliders there is no doubt that the required energy density $\gtrsim 2 ~{\rm GeV}/{\rm fm}^3$ to form the quark-gluon plasma at RHIC and LHC is achieved.

However, there is no proof that the quark-gluon plasma formed at RHIC and LHC is thermalized. This is due to the following reason. At the center of mass energy per nucleon ${\sqrt s}_{NN}$ = 200 GeV at RHIC and at the center of mass energy per nucleon ${\sqrt s}_{NN}$ = 5.02 TeV at LHC the two heavy nuclei before the collision travel almost at the speed of light. Since the two nuclei at RHIC and LHC travel almost at the speed of light the longitudinal momenta of the partons inside the nuclei are much larger than their transverse momenta. This leads to large momentum anisotropy of partons just after the nuclear collisions at RHIC and LHC. Hence we find that the quark-gluon plasma formed at RHIC and LHC at the initial time is in non-equilibrium.

This means we know for sure that the quark-gluon plasma formed at the initial time at RHIC and LHC is in non-equilibrium but we do not know for sure that this non-equilibrium quark-gluon plasma is thermalized at RHIC and LHC. In order for the non-equilibrium quark-gluon plasma at RHIC and LHC to thermalize many more secondary partonic collisions are necessary over a long period of time. However, there is no way of directly determining the time period over which these secondary partonic collisions in non-equilibrium take place. This is because we do not directly experimentally measure quarks and gluons but we directly experimentally measure hadrons and other color singlet non-hadronic observables. All we know is that the hadronization time scale in QCD is very small $\sim ~1 fm/c~\sim10^{-24}$ seconds.

Since the hadronization time scale in QCD is very small ($\sim ~1 fm/c~\sim10^{-24}$ seconds) and since the two nuclei travel almost at the speed of light at RHIC and LHC, it is unlikely that there are many more secondary partonic collisions over a long period of time to bring this non-equilibrium quark-gluon plasma to equilibrium.

For this reason one should not assume that a thermalized quark-gluon plasma is formed at RHIC and LHC even if an assumption of thermalized quark-gluon plasma at a later time describes the experimental data. In order to prove that the quark-gluon plasma is thermalized at RHIC and LHC one must prove that the same experimental data can not be described by the non-equilibrium quark-gluon plasma. This is because we know for sure that the quark-gluon plasma is formed in non-equilibrium just after the nuclear collisions at RHIC and LHC.

In this paper we describe non-equilibrium QCD by using closed-time path integral formalism and show that the exact first principle method to study quark-gluon plasma at RHIC and LHC is the nonequilibrium-nonperturbative QCD by using the closed-time path integral formalism.

Hence to prove that the same experimental data at RHIC and LHC can not be described by the non-equilibrium quark-gluon plasma from the first principle one has to study nonequilibrium-nonperturbative QCD using closed-time path integral formalism. The nonperturbative QCD is necessary at RHIC and LHC for three purposes: 1) to study the soft parton production in initial nuclear collisions which plays a major role to determine the bulk properties of the quark-gluon plasma, 2) to calculate the cross section of the secondary partonic collisions of soft partons in non-equilibrium which can not be calculated by using pQCD, and 3) to study how the partons from the QGP form hadrons at RHIC and LHC heavy-ion colliders.

Hence we conclude that in the absence of the first principle nonequilibrium-nonperturbative QCD study by using the closed-time path integral formalism at RHIC and LHC there is no proof of thermalized quark-gluon plasma at RHIC and LHC heavy-ion colliders from the first principle.

The paper is organized as follows. In section II we discuss the limitation of the lattice QCD at finite temperature at RHIC and LHC. In section III we discuss the limitation of the hydrodynamics at RHIC and LHC. In section IV we show that there is no proof of color glass condensate at RHIC and LHC heavy-ion colliders from the first principle. In section V we show that AdS/CFT and super Yang-Mills (SYM) plasma are not valid at RHIC and LHC. In section VI we discuss the major drawback of the jet quenching study at RHIC and LHC in the literature. In section VII we discuss the non-equilibrium QCD using the closed-time path integral formalism and show that there is no proof of thermalized quark-gluon plasma at RHIC and LHC from the first principle. Section VIII contains conclusions.

\section{ Lattice QCD at finite temperature is not applicable for non-equilibrium quark-gluon plasma at RHIC and LHC }

There is lot of activities on lattice QCD at finite temperature \cite{l1,l2,l3,l4}. The primary motivation of such large scale computer simulation is to study the properties of the quark-gluon plasma formed at the high energy heavy-ion colliders at RHIC and LHC. However, as we have discussed above, there is no proof of thermalized quark-gluon plasma at RHIC and LHC. If the quark-gluon plasma is not thermalized then one cannot even define a temperature. If one cannot define a temperature then one cannot study the lattice QCD at finite temperature. Since there is no proof of thermalized quark-gluon plasma at RHIC and LHC, there is no proof that the quantities predicted by the lattice QCD simulations at finite temperature \cite{l1,l2,l3,l4} are of any relevance to the actual heavy-ion collisions RHIC and LHC.

We mention again here that the lattice QCD at finite temperature is not applicable in non-equilibrium QCD because one can not define a temperature $T$ in non-equilibrium. Hence if the quark-gluon plasma at RHIC and LHC is in non-equilibrium from beginning to end (from initial time to hadronization) then the lattice QCD at finite temperature is not applicable at RHIC and LHC at all.

Hence, as far as the actual science at the RHIC and LHC heavy-ion colliders is concerned, it is not desirable to put lot of efforts on lattice QCD simulation at finite temperature. What one needs is to study the nonequilibrium-nonperturbative QCD by using closed-time path integral formalism which is the exact first principle method to study the quark-gluon plasma at RHIC and LHC. Once one knows how to study nonequilibrium-nonperturbative QCD using closed-time path integral formalism then one can easily study the thermalized quark-gluon plasma from it.

Hence one of the major challenge for us at RHIC and LHC heavy-ion colliders is to study nonequilibrium-nonperturbative QCD by using closed-time path integral formalism. The nonequilibrium-nonperturbative QCD by using the closed-time path integral formulation is one of the frontier area of research in QCD due to the experiments at the RHIC and LHC heavy-ion colliders.

\section{ Hydrodynamics is not applicable for non-equilibrium quark-gluon plasma at RHIC and LHC }

Hydrodynamics is widely applied to study quark-gluon plasma at RHIC and LHC heavy-ion colliders \cite{s1,s2,s3,s4,s5,s6,s7}. In fact the conclusion of the thermalized  quark-gluon plasma reached by the heavy-ion community is primarily based on the hydrodynamics study. In the hydrodynamics study one assumes that the quark-gluon plasma is thermalized at certain time $t_0$ which is then fitted from the experimental data. This is a very serious oversimplification of the one of the most complicated collisions in QCD at the RHIC and LHC heavy-ion colliders.

Note that the QCD at high energy $pp$ colliders is already very complicated. Hence the QCD at high energy $AuAu$ colliders at RHIC and at high energy $PbPb$ colliders at LHC are much more complicated than the QCD at high energy $pp$ colliders. This implies that there is no justification to study hydrodynamics assuming thermalization \cite{s1,s2,s3,s4,s5,s6,s7} at RHIC and LHC instead of studying QCD at high energy $AuAu$ colliders at RHIC and at high energy $PbPb$ colliders at LHC.

As discusse above, the quark-gluon plasma at the initial time is formed in non-equilibrium where the hydrodynamics is not applicable. If the quark-gluon plasma is in non-equilibrium from beginning to end (from initial time to hadronization) then there is no room for the application of hydrodynamics at the RHIC and LHC heavy-ion colliders. Another limitation of hydrodynamics is that it can not implement how an identified hadron is formed from partons from the quark-gluon plasma as it requires nonperturbative QCD.

One may argue that since the experimental data at RHIC and LHC heavy-ion colliders are fitted with hydrodynamics assuming thermalization, the quark-gluon plasma at RHIC and LHC is thermalized \cite{s1,s2,s3,s4,s5,s6,s7}. However, this argument is not correct unless one proves that the same experimental data at RHIC and LHC heavy-ion colliders can not be explained by the non-equilibrium quark-gluon plasma. In order to prove that the same experimental data at RHIC and LHC heavy-ion colliders can not be explained by the non-equilibrium quark-gluon plasma from the first principle one has to study nonequilibrium-nonperturbative QCD by using closed-time path integral formalism.

\section{ There Is No proof of color glass condensate at RHIC and LHC heavy-ion colliders from the first principle }

The color glass condensate \cite{m1,m2} was initially designed to estimate the parton distribution at small transverse momentum for very large nucleus at very high energy. First of all, the QCD is the first principle theory to describe quarks and gluons. The hard parton production can be calculated from the first principle by using perturbative QCD (pQCD) which is well known. The soft parton production from the first principle can be calculated by using non-perturbative QCD which is not known yet. Since the non-perturbative QCD is not solved yet, any effective theory to compute parton distribution function for large nucleus is not a first principle calculation.

In addition to this there are further complications at RHIC and LHC heavy-ion colliders due to the formation of non-equilibrium quark-gluon plasma. Since we measure the final state hadrons (and other final state color singlet non-hadronic observables) at RHIC and LHC we do not directly measure the initial condition of the parton distribution function for large nucleus. Hence in order to prove color glass condensate at RHIC and LHC using the hadronic (or other color singlet non-hadronic observables) from the first principle one has to study nonequilibrium-nonperturbative QCD by using the closed-time path integral formalism. In the absence of the nonequilibrium-nonperturbative QCD calculation by using closed-time path integral formalism there is no proof of color glass condensate at RHIC and LHC heavy-ion colliders from the first principle.

\section{ AdS/CFT and Super Yang-Mills plasma are not valid at RHIC and LHC heavy-ion colliders }

As mentioned above the analytical solution of the non-perturbative QCD is not available yet. In recent years there has been lot of efforts in the study of AdS/CFT \cite{r1,r2} and the super Yang-Mills (SYM) plasma \cite{sp1} to understand the non-perturbative aspects of the quark-gluon plasma. However, it should be mentioned here that the study of plasma based on the AdS/CFT calculation is not the same as the study of the actual quark-gluon plasma in QCD. Similarly the study of the super Yang-Mills plasma is not same as the actual quark-gluon plasma in QCD. Hence the determination of any quantities based on the AdS/CFT and SYM plasma scenario do not tell us anything about the corresponding quantities in the actual quark-gluon plasma in QCD.

In addition to this the AdS/CFT is not valid at the RHIC and LHC heavy colliders experiments because string theory is not experimentally verified and we have not found any experimental evidence of TeV scale string theory at LHC.

Similarly the SYM plasma is not valid at the RHIC and LHC heavy colliders experiments because the supersymmetry is not experimentally verified and we have not found any experimental evidence of TeV scale supersymmetry at LHC.

\section{ Major Drawback of the jet quenching study at RHIC and LHC heavy-ion colliders }

We have observed back-to-back jets in $pp$ collisions at high energy colliders. However, in heavy-ion colliders at RHIC and LHC we have not observed back-to-back jets. This is reported by the STAR \cite{j1} and PHENIX \cite{j2} collaborations at RHIC in Au-Au collisions and by ATLAS \cite{j3}, CMS \cite{j4} and ALICE \cite{j5} collaborations at LHC in Pb-Pb collisions which is known as the jet quenching. It is expected that the jet quenching is due to the presence of quark-gluon plasma medium where one of the back-to-back jet is resolved or completely destroyed.

The back-to-back jets in $pp$ collisions is studied by using QCD in vacuum. However, the QCD in vacuum is not applicable in Au-Au collisions at RHIC and in Pb-Pb collisions at LHC. This is because the initial state at RHIC and LHC heavy-ion colliders is not the vacuum state $|0>$ any more due to the presence of the QCD medium. Instead, the initial state at RHIC and LHC heavy-ion colliders is $|in>$ which contains the information of both the vacuum and the medium at the initial time.

In the back-to-back jets study in QCD in vacuum the definition of the parton to hadron fragmentation function is given Collins and Soper in \cite{jv}.

Recently we have derived the definition of the parton to hadron fragmentation function in non-equilibrium QCD by using closed-time path integral formalism \cite{qf1}. The definition of the quark to hadron fragmentation function in non-equilibrium QCD is given by \cite{qf1,qf}
\bea
&& D_q(z,k_T) = \frac{1}{12z[1-f_q(p^+,p_T)]} \int dy^- \int \frac{d^2y_T}{(2\pi)^3} e^{ip^+y^-+ik_T \cdot y_T/z} \nonumber \\
&& {\rm Tr}\{\gamma^+<in|\psi(y^-,y_T)W[y^-,y_T]a^\dagger_H(k^+,0_T) a_H(k^+,0_T) W[0]\psi(0)|in>\}
\label{qfn}
\eea
where $\psi(x)$ is the quark field, $z$ is the longitudinal momentum fraction of the hadron with respect to the parton, $p^\mu$ is momentum of the parton, $k^\mu$ is the momentum of the hadron, $f_q({\vec p})$ is the non-equilibrium distribution function of the quark and the light-like Wilson line $W[y]$ is given by
\bea
W[y]={\cal P}e^{-igT^b\int_0^\infty d\lambda l \cdot A^b(y+l\lambda)}
\label{lw}
\eea
where $l^\mu$ is the light-like four-velocity and $A_\nu^b(x)$ is the SU(3) pure gauge background field.

Similarly the definition of the gluon to hadron fragmentation function in non-equilibrium QCD is given by \cite{gf}
\bea
&& D_g(z,k_T) = \frac{p^+}{16z[1+f_g(p^+,p_T)]} \int dy^- \int \frac{d^2y_T}{(2\pi)^3} e^{ip^+y^-+ik_T \cdot y_T/z} \nonumber \\
&& <in|Q^{\nu b}(y^-,y_T)W^{(A)}[y^-,y_T]a^\dagger_H(k^+,0_T) a_H(k^+,0_T) W^{(A)}[0]Q_\nu^b(0)|in>
\label{gfn}
\eea
where $Q_\nu^b(x)$ is the (quantum) gluon field, $f_g({\vec p})$ is the non-equilibrium distribution function of the gluon and the light-like Wilson line $W^{(A)}[y]$ in the adjoint representation of SU(3) is given by
\bea
W^{(A)}[y]={\cal P}e^{-igT^b_{adj}\int_0^\infty d\lambda l \cdot A^b(y+l\lambda)},~~~~~~~~~~~~~~~T^{b}_{cd,~adj}=-if^{bcd}.
\label{lwa}
\eea
Note that when $|in>=|0>$ we have $f({\vec p})=0$ and hence the eqs. (\ref{qfn}) and (\ref{gfn}) reproduce the fragmentation functions in vacuum. However, in non-equilibrium QCD the initial state $|in>$ contains the information of both the vacuum and the medium at the initial time. But since the parton to hadron fragmentation functions in eqs. (\ref{qfn}) and (\ref{gfn}) are non-perturbative quantities in QCD one finds that it is not possible to separate the vacuum part of the fragmentation from the medium part of the fragmentation function.

This implies that one can not use the fragmentation function in vacuum (for example from $e^+e^-$ and $pp$ colliders) to study the hadron production from the quark-gluon plasma at RHIC and LHC.

Since the jet quenching studies in the literature have directly/indirectly used the value of the fragmentation function from vacuum (for example from $e^+e^-$ and $pp$ colliders) we find that the jet quenching studies in the literature \cite{iv1,iv2} at the RHIC and LHC heavy-ion colliders are not correct from the first principle.

\section{ There Is No Proof of Thermalized Quark-Gluon Plasma at RHIC and LHC From First Principle }

As described earlier the ground state at the initial time at the RHIC and LHC heavy-ion colliders is given by $|in>$ which is not the vacuum state $|0>$ because of the presence of the medium. This ground state $|in>$ at the initial time contains the information of the vacuum and the medium. For example when the initial distribution function $f(p)=0$ then the state $|in>$ reproduces to vacuum state $|0>$.

Since the gluons in QCD are self interacting the path integral formalism is necessary to study QCD. The non-equilibrium quantum field theory can be studied by using closed-time path formalism. Although canonical quantization formalism can be used to study non-equilibrium scalar field theory and non-equilibrium QED but it is not suitable to study the non-equilibrium QCD due to self interacting gluons. For this reason we study the non-equilibrium QCD by using the closed-time path integral formulation. 

\subsection{Non-Equilibrium QCD at RHIC and LHC using closed-time path integral formalism }

The generating functional (including heavy quark) in the non-equilibrium QCD by using closed-time path integral formalism is given by
\bea
&&Z[K,J^g_+,J^g_-,\eta^u_+,\eta^u_-,{\bar \eta}^u_+,{\bar \eta}^u_-,\eta^d_+,\eta^d_-,{\bar \eta}^d_+,{\bar \eta}^d_-,\eta^s_+,\eta^s_-,{\bar \eta}^s_+,{\bar \eta}^s_-,\eta^H_+,\eta^H_-,{\bar \eta}^H_+,{\bar \eta}^H_-]\nonumber \\
&&=\int [dG_+][dG_-][d\psi_u][d{\bar \psi}_u][d\psi_s][d{\bar \psi}_d][d\psi_s][d{\bar \psi}_s][d\psi_H][d{\bar \psi}_H]~\times ~{\rm det}[\frac{\delta \partial^\nu G^d_{\nu +}}{\delta \omega^a_+}]~\times ~{\rm det}[\frac{\delta \partial^\nu G^d_{\nu -}}{\delta \omega^a_-}]\nonumber \\
&& \times~ {\rm exp}[i\int d^4x [-\frac{1}{4}F_{\mu \lambda}^{h}[G_+]F^{\mu \lambda h}[G_+] +\frac{1}{4}F_{\mu \lambda}^{h}[G_-]F^{\mu \lambda h}[G_-]-\frac{1}{2\beta }(\partial^\mu G_{\mu +}^c)^2 +\frac{1}{2\beta }(\partial^\mu G_{\mu -}^c)^2 \nonumber \\
&& +{\bar \psi}_{u+}[i {\not \partial}-m_u+gT^c{\not G}^c_+]\psi_{u+} -{\bar \psi}_{u-}[i {\not \partial}-m_u+gT^c{\not G}^c_-]\psi_{u-} \nonumber \\
&& +{\bar \psi}_{d+}[i {\not \partial}-m_d+gT^c{\not G}^c_+]\psi_{d+} -{\bar \psi}_{d-}[i {\not \partial}-m_d+gT^c{\not G}^c_-]\psi_{d-} \nonumber \\
&& +{\bar \psi}_{s+}[i {\not \partial}-m_s+gT^c{\not G}^c_+]\psi_{s+} -{\bar \psi}_{s-}[i {\not \partial}-m_s+gT^c{\not G}^c_-]\psi_{s-}\nonumber \\
&& +{\bar \psi}_{H+}[i {\not \partial}-m_H+gT^c{\not G}^c_+]\psi_{H+} -{\bar \psi}_{H-}[i {\not \partial}-m_H+gT^c{\not G}^c_-]\psi_{H-}+ J^g_+ \cdot G_+ - J^g_- \cdot G_- \nonumber \\
&& + {\bar \eta}_{u+} \psi_{u+}  + {\bar \psi}_{u+} \eta_{u+} - {\bar \eta}_{u-} \psi_{u-}  - {\bar \psi}_{u-} \eta_{u-}+ {\bar \eta}_{d+} \psi_{d+}  + {\bar \psi}_{d+} \eta_{d+} - {\bar \eta}_{d-} \psi_{d-}  - {\bar \psi}_{d-} \eta_{d-}  \nonumber \\
&& + {\bar \eta}_{s+} \psi_{s+}  + {\bar \psi}_{s+} \eta_{s+} - {\bar \eta}_{s-} \psi_{s-}  - {\bar \psi}_{s-} \eta_{s-}+ {\bar \eta}_{H+} \psi_{H+}  + {\bar \psi}_{H+} \eta_{H+} - {\bar \eta}_{H-} \psi_{H-}  - {\bar \psi}_{H-} \eta_{H-} ]] \nonumber \\
&&\times ~<G_+,\psi_{u+},{\bar \psi}_{u+},\psi_{d+},{\bar \psi}_{d+},\psi_{s+},{\bar \psi}_{s+},\psi_{H+},{\bar \psi}_{H+},0|K|0,{\bar \psi}_{H-},\psi_{H-},{\bar \psi}_{s-},\psi_{s-},{\bar \psi}_{d-},\psi_{d-}\nonumber \\
&&,{\bar \psi}_{u-}, \psi_{u-}, G_->
\label{zfqd}
\eea
where $\pm$ indices are the closed-time path indices, $K$ is the initial density of states, $G_\mu^a(x)$ is the (quantum) gluon field, $\psi(x)$ is the quark field, $u,d,s$ are the light quarks, $H$ represents heavy quark, $\beta$ is the gauge fixing parameter, $J_\mu^{ag}(x)$ is the source to the gluon field, ${\bar \eta}(x)$ is the source to the quark field and
\bea
F_{\mu \lambda}^{h}[G] = \partial_\mu G_\lambda^h(x) - \partial_\lambda G_\mu^h(x) +gf^{hbd} G_\mu^b(x)\partial_\mu G_\lambda^d(x).
\label{fml}
\eea
Note that we do not introduce ghost field in eq. (\ref{zfqd}) because we directly work with the ghost determinant ${\rm det}[\frac{\delta \partial^\nu G^d_{\nu}}{\delta \omega^a}]$ in this paper. Since we work in the frozen ghost formalism at the initial time \cite{kg,kg1} we do not have any ghost fields in the kernel $<G_+,\psi_{u+},{\bar \psi}_{u+},\psi_{d+},{\bar \psi}_{d+},\psi_{s+},{\bar \psi}_{s+},\psi_{H+},{\bar \psi}_{H+},0|K|0,{\bar \psi}_{H-},\psi_{H-},{\bar \psi}_{s-},\psi_{s-},{\bar \psi}_{d-},\psi_{d-},{\bar \psi}_{u-}, \psi_{u-}, G_->$ in eq. (\ref{zfqd}).

As can be seen from eq. (\ref{zfqd}) it is extremely difficult (sometimes impossible) to calculate a physical quantity by using nonequilibrium-nonperturbative QCD at RHIC and LHC in the closed-time path integral formalism. However, this is the correct first principle method to study the quark-gluon plasma at RHIC and LHC. Note that even if the generating functional in eq. (\ref{zfqd}) in non-equilibrium QCD is extremely difficult to calculate many physical quantities at RHIC and LHC but some of the physics can be exactly studied at RHIC and LHC by using eq. (\ref{zfqd}).

For example, recently we have used the generating functional $Z[K,J^g_+,J^g_-,\eta_+,\eta_-,{\bar \eta}_+,{\bar \eta}_-]$ in eq. (\ref{zfqd}) to derive the gauge invariant quark and gluon to hadron fragmentation functions in non-equilibrium QCD as given by eq. (\ref{qfn}) and (\ref{gfn}) which are consistent with the factorization of infrared divergences at all orders in coupling constant. The factorization theorem plays an important role in QCD at high energy colliders \cite{alfa,alfa1}.

Similarly the generating functional $Z[K,J^g_+,J^g_-,\eta^u_+,\eta^u_-,{\bar \eta}^u_+,{\bar \eta}^u_-,\eta^d_+,\eta^d_-,{\bar \eta}^d_+,{\bar \eta}^d_-,\eta^s_+,\eta^s_-,{\bar \eta}^s_+,{\bar \eta}^s_-$\\
$,\eta^H_+,\eta^H_-,{\bar \eta}^H_+,{\bar \eta}^H_-]$ including heavy quark $H$ in eq. (\ref{zfqd}) in non-equilibrium QCD by using the closed-time path integral formalism is used to prove the factorization of $\chi_{cJ}$ and NRQCD heavy quarkonium production from the non-equilibrium quark-gluon plasma at RHIC and LHC \cite{nch,nn}. 

\subsection{Nonequilibrium-Nonperturbative QCD using closed-time path integral formalism is the Exact First Principle Method to Study Quark-Gluon Plasma at RHIC and LHC}

As mentioned earlier, since the two nuclei at RHIC and LHC travel almost at the speed of light the longitudinal momenta of the partons inside the nuclei are much larger than their transverse momenta leading to non-equilibrium quark-gluon plasma formation at the initial time just after the nuclear collisions at RHIC and LHC \cite{qn5,qn6,qn7,qn8}. In addition to the two nuclei moving almost at the speed of light, since the hadronization time scale in QCD is very small ($\sim ~1 fm/c~\sim10^{-24}$ seconds) there may not be enough time for many more secondary partonic collisions to take place to form the thermalized quark-gluon plasma at RHIC and LHC.

One may argue that the quark-gluon plasma is thermalized because the experimental data at RHIC and LHC heavy-ion colliders are described by hydrodynamics if one assumes thermalization at a later time \cite{s1,s2,s3,s4,s5,s6,s7}. However, this argument is not correct because in order to make sure that the thermalized quark-gluon plasma is formed at RHIC and LHC one has to prove that the same experimental data at RHIC and LHC can not be explained by the non-equilibrium quark-gluon plasma. This is because we know for sure that the quark-gluon plasma is formed in non-equilibrium just after the nuclear collisions. If the quark-gluon plasma is in non-equilibrium from beginning to end (from initial time to hadronization) then the hydrodynamics is not applicable at RHIC and LHC.

Closed-time path (CTP) formalism is the first principle method to study non-equilibrium quantum field theory. The non-equilibrium QCD can be studied by using closed-time path integral formulation because of the self interacting gluons.

As mentioned earlier the nonperturbative QCD is necessary at RHIC and LHC for three reasons: 1) to study the soft parton production in initial nuclear collisions which plays a major role to determine the bulk properties of the quark-gluon plasma, 2) to calculate the cross section of the secondary partonic collisions of soft partons in non-equilibrium which can not be calculated by using pQCD (these soft partonic scattering cross section in non-equilibrium plays an important role to study thermalization), and 3) to study how the final state hadrons are formed from the partons from the quark-gluon plasma at RHIC and LHC heavy-ion colliders.

Hence one finds that the nonequilibrium-nonperturbative QCD by using the closed-time path integral formalism is the correct first principle method to study the quark-gluon plasma at RHIC and LHC.

For this reason one finds that there is no proof of thermalized quark-gluon plasma at RHIC and LHC from the first principle.

\section{Conclusions}
Although Tevatron has discovered top quark and LHC (pp collisions) has discovered Higgs boson but the RHIC and LHC heavy-ion colliders have not discovered thermalized quark-gluon plasma (QGP). This is because the experimental data of top quark at Tevatron and the experimental data of Higgs boson at LHC are compared with the exact first principle calculation but the experimental data at RHIC and LHC are compared with simplistic models and assumptions which are not exact first principle calculation. In this paper we have shown that the exact first principle method to study quark-gluon plasma at RHIC and LHC is the nonequilibrium-nonperturbative QCD by using closed-time path integral formalism. Hence in the absence of such exact first principle calculation we have concluded that there is no proof of thermalized quark-gluon plasma at RHIC and LHC.

\end{document}